\documentclass[preprintnumbers,showpacs,showkeys,preprint,secnumarabic,amssymb,amsmath,amsfonts,aps, pre]{revtex4-1}

\usepackage{calrsfs}
\DeclareMathAlphabet{\pazocal}{OMS}{zplm}{m}{n}

\usepackage{makeidx}
\usepackage{gensymb}
\usepackage{amsfonts}
\usepackage{ragged2e}
\usepackage[pdftex]{graphicx, color}
\usepackage[dvipsnames]{xcolor}
\usepackage{float}
\usepackage{here}
\usepackage{fullpage}
\usepackage{color}

\newcommand{\pp}{\mathcal{P}}
\newcommand{\D}{\mathrm{d}}
\usepackage{epsfig}
\definecolor{darkWhite}{rgb}{0.94,0.94,0.94}

\begin{document}
%%%%%%%%%%%%%%%%%%%%%%%%%%%%%%%%%%%%%%%%%%%%%%%%%%%%%%%%%
% AUTHOR INFORMATIONS

\title{High order derivatives of Boltzmann microcanonical entropy with an additional conserved quantity}

\date{\today}

\author{Ghofrane Bel-Hadj-Aissa}
\email{ghofrane.belhadjaissa@gmail.com}
\affiliation{Physics Department, University of Siena, Via Roma 56, 53100 Siena, Italy}

%%%%%%%%%%%%%%%%%%%%%%%%%%%%%%%%%%%%%%%%%%%%%%%%%%%%%%%%
\begin{abstract}
 In this article, using a known method \cite{PearsonLaplace1985}, a computation is performed of the derivatives of the microcanonical entropy, with respect to the energy up to the 4-th order, using a Laplace transform technique,  and adapted it to the case where the total momentum is conserved. The outcome of this computation answers a theoretical question concerning the description of thermodynamics associated with a Hamiltonian flow in presence of an additional conserved quantity besides energy.
 This is also  of practical interest in numerical simulations of the microcanonical thermodynamics associated to classical Hamiltonian flows.
\end{abstract}

% BODY OF PAPER
%%%%%%%%%%%%%%%%%%%%%%%%%%%%%%%%%%%%%%%%%%%%%%%%%%%%%%%%%
\maketitle

\section{Introduction}
After the Poincaré-Fermi theorem \cite{poincare189299,fermi1,fermi2}, a generic property of non linear Hamiltonian systems is that the only non trivial conserved quantity is the total energy of the system, this is true for Hamiltonians which do not explicitly depend on time. However, invariance under spatial translations and rotations add conserved total momentum and angular momentum, respectively. Since the invariant measure in phase space for a Hamiltonian flow is the microcanonical measure, Hamiltonian dynamics can be used to compute thermodynamic quantities \cite{PearsonLaplace1985,cerruti2001clustering,pettini2007geometry,Franzosi}. Therefore, we can wonder how can we describe the thermodynamics of systems with the above mentioned additional conserved quantities, because these constrain the Hamiltonian flow on hypersurfaces of phase space of co-dimension larger than one. 
\\ Moreover, through Hamiltonian flows, one can study phase transitions in the microcanonical ensemble \cite{pettini2007geometry}, and, in this framework,  Qi and Bachmann proposed \cite{qi2018classification} a definition for a k-th order phase transitions in the microcanonical ensemble through the analysis of the inflection points of the  entropy and its derivatives. This is what particularly motivates the present investigation of how and to what extent the already known \cite{PearsonLaplace1985} analytic expressions of  thermodynamic observables have to be corrected when an additional global invariant is taken into account, and a numerical study of phase transitions through the associated Hamiltonian flow is undertaken.

\section{Computing the derivatives of the entropy}
The outcomes of the computations reported throughout this work apply to systems described by standard Hamiltonian functions, that is
\begin{equation}
H(p^{1},..p^{N},q^{1},...,q^{N})=\sum_{i=1}^{N} \dfrac{(p^{i})^2}{2}+V(q^{1},...,q^{N})\;,
\end{equation}
where $N$ is the number of degrees of freedom, $p$ is the momentum and $V$ is the potential energy, invariant for the global translation $V(q_1+a,...,q_N+a)=V(q_1,...,q_N)$.
\\According to Noether's theorem, such system has two first integrals of motion: the total energy $H=E$ and the total momentum $P=\sum_{i=1}^N p^i$.
\\ As a consequence, the Hamiltonian flow is constrained on co-dimension-two subspaces of the phase space $\Lambda$, that is
\begin{equation}
\Sigma_{E,\pp}=\{x=(p^{1},..p^{N},q^{1},...,q^{N})\in\Lambda \, \,|\,\,\, P(x)=\pp \, \text{and} \,H(x)=E\}\,\,.
\end{equation}
It follows that the microcanonical partition function according to the Boltzmann prescription reads
\begin{equation}
\Omega(E,\pp)=\int_{\Lambda}\delta\left(H(p,q)-E\right)\delta\left(P-\pp\right)\,\,\,\prod_{i=1}^{N}\D p^i\wedge \D q^{i}
\end{equation}
so the specific entropy is
\begin{equation}
\label{eq:microcan_spec_S}
S(\varepsilon,\pp)=\dfrac{1}{N}\ln \left(\dfrac{\Omega(N \varepsilon,\pp)}{\Omega_{0}}\right),
\end{equation}

where $\Omega_{0}$ is an arbitrary constant, and $\varepsilon=E/N$ is the specific energy.
\\In what follows we choose to fix $\Omega_{0}=1$ and set for simplicity, but without loss of generality, the total momentum $P$ equal to zero, that is $\pp =0$. To simplify the notation, we omit the total momentum fixed to zero in both the argument of microcanonical entropy
$S(\varepsilon,0)\rightarrow S(\varepsilon)$ and of the microcanonical partition function $\Omega(E,0)\rightarrow\Omega(E)$.
All the relevant thermodynamic observables, such as temperature or specific heat, can be expressed as a function 
of the derivatives of the specific entropy $S$ with respect to the  specific energy\footnote{In the microcanonical ensemble, all the thermodynamic observables can be obtained by deriving the entropy with respect to other state variables, e.g. the pressure $p=T (\partial_{v} S)_{\varepsilon}$ depends on the derivative of the entropy with respect to the volume $v$. In what follows, we do not consider the dependence of the microcanonical entropy on other state variables but the specific energy $\varepsilon=E/N$.} $\varepsilon$.
The derivatives of the specific entropy with respect to the specific energy up to the fourth order read
\begin{equation}
\label{eq:derEnt_microcan}
\begin{split}
&\dfrac{\partial S}{\partial\varepsilon}=\dfrac{\Omega^{'}}{\Omega}\\
&\dfrac{\partial^{2} S}{\partial\varepsilon^{2}} =N\bigg[\dfrac{\Omega^{''}}{\Omega}-\bigg (\dfrac{\Omega^{'}}{\Omega}\bigg)^{2}\bigg]\\
&\dfrac{\partial^{3} S}{\partial\varepsilon^{3}}=N^{2}\left[\dfrac{\Omega^{'''}}{\Omega}-3\dfrac{\Omega^{''}}{\Omega}\dfrac{\Omega^{'}}{\Omega}+2\left(\dfrac{\Omega^{'}}{\Omega}\right)^{3}\right]\\
&\dfrac{\partial^{4} S}{\partial\varepsilon^{4}}=N^{3}\left[\dfrac{\Omega^{''''}}{\Omega}-4\dfrac{\Omega^{'''}}{\Omega}\,\dfrac{\Omega^{'}}{\Omega}-3\left(\dfrac{\Omega^{''}}{\Omega}\right)^{2}+12\dfrac{\Omega^{''}}{\Omega}\left(\dfrac{\Omega^{'}_N}{\Omega}\right)^2-6\left(\dfrac{\Omega^{'}}{\Omega}\right)^{4}\right]\;,
\end{split}
\end{equation}
where the prime corresponds to the derivative with respect to the total energy $E$. 
In order to characterize the microcanonical thermodynamics of a given system, a method is needed allowing to calculate the higher order derivatives of microcanonical entropy \eqref{eq:derEnt_microcan}.\\
In Ref.\cite{PearsonLaplace1985} a method is presented that allows to derive the expressions of thermodynamic observables in terms of the average of the (specific) kinetic energy and its powers on the $\Sigma_{E}$ when the \textit{only constraint is the fixed total energy of the system} $H=E$ and for systems whose Hamiltonian is of the form
\begin{equation}
H(p^{1},..p^{N},q^{1},...,q^{N})=K(p^{1},..p^{N})+V(q^{1},...,q^{N})\;,
\end{equation}
where $K$ is the total kinetic energy.
Such a method is based on a Laplace transform technique applied to the microcanonical partition function $\Omega$ and it allows performing integration on the $p^i$ variables.\\
In this letter, we show how to further develop and apply this technique to the case where the total momentum $P$ of the system  is conserved and set equal to zero.
Let us consider the Laplace transform of the microcanonical partition function $\Omega$
\begin{equation}
\begin{split}
\mathcal{L}[\Omega(t)]&=\int_{0}^{+\infty}\, e^{-tE}\,\Omega(E)\,\D E\\
&=\int \prod^{N}_{i=1} \D p_i\,\delta (P)\int\, \prod^{N}_{i=1} \D q_i\, e^{-t H(\{p^{1},\ldots,p^{N},q^{1},\ldots,q^{N}\})}\\
&=\int \prod^{N}_{i=1} \D p_i\, e^{-t \sum_{i=1}^{N}\frac{(p^{i})^2}{2}}\,\delta(P)\, \int \prod^{N}_{i=1} \D q_i e^{-tV(q^{1},\ldots,q^{N})}\;.
\end{split}
\end{equation}
In Ref.\cite{PearsonLaplace1985} the integration over the $N$ momenta $p^i$ is easily performed, as it is a Gaussian integral. In the case here considered, however, we have to take into account also the constraint on $P$. In order to do this, we consider the following identity for the Dirac delta function
\begin{equation}
\delta(P)=\dfrac{1}{2\pi}\int_{-\infty}^{+\infty} \D s\, e^{i s P}\;,
\end{equation}
so that
\begin{equation}
\begin{split}
\mathcal{L}[\Omega](t)&=\dfrac{1}{2\pi}\int \prod^{N}_{i=1} \D p_i\,e^{-t\sum_{i=1}^{N}\frac{(p^{i})^2}{2}}\,\int_{-\infty}^{+\infty} \D s\, e^{i s P}\,\int \prod^{N}_{i=1} \D q_i e^{-tV(q^{1},\ldots,q^{N})}\\
&=\dfrac{1}{2\pi}\int \prod^{N}_{i=1} \D p_i\,\int_{-\infty}^{+\infty}\,\D s \, e^{-\frac{t}{2}\sum_k (p^k)^2}\,\,e^{is\sum_k p_k}\,\int \prod^{N}_{i=1} \D sq_i e^{-tV(q^{1},\ldots,q^{N})}\\
&=\dfrac{1}{2\pi}\int \prod^{N}_{i=1} \D p_i\, \int_{-\infty}^{+\infty} \D s \, e^{-\frac{t}{2}\sum_k (p_k -is/t)^2}\,e^{-\frac{Ns^2}{2t}}\,\int \prod^{N}_{i=1} \D q_i e^{-tV(q^{1},\ldots,q^{N})}\\
&=\dfrac{1}{(2\pi)^{\left(1-\frac{N}{2}\right)}}\, \int_{-\infty}^{+\infty} \D s\, e^{-\frac{Ns^2}{2t}}\,t^{-N/2} \, \int \prod^{N}_{i=1} \D q_i e^{-tV(q^{1},\ldots,q^{N})}\\
&=\underbrace{\frac{1}{\sqrt{N}}\,\dfrac{1}{(2\pi)^{\frac{1}{2}\left(1-N\right)}}}_{1/C}\, \int \prod^{N}_{i=1} \D q_i \, t^{-(N-1)/2}\,e^{-tV(q^{1},\ldots,q^{N})}\,\,\,.
\end{split}
\end{equation}
We now use the Bromwich integral to inverse the Laplace transform
\begin{equation*}
\Omega (E) =\dfrac{1}{2\pi i \,C} \int_{\gamma -i\infty} ^{\gamma +i\infty} \int \prod^{N}_{i=1} \D q_i\, t^{-(N-1)/2}\, e^{t\left[E-V\left(q^{1}\ldots q^{N}\right)\right]} \, \D t\;,
\end{equation*}
where $\gamma$ is a vertical contour in the complex plane chosen so that all the singularities of $t^{-(N-1)/2}$ are on the left of the vertical part of the integration contour. 

After using the theorem of residues, one obtains the final expression for $\Omega$ in the form of an integral over the
manifold $M_{E}=\left\{(q^1,...,q^N)\in\Lambda_q\,|\, V(q^1,\ldots,q^N)\leq E\right\}$ as 
\begin{equation}
\Omega(E)=\underbrace{\dfrac{1}{ C \,\Gamma \left(\dfrac{N}{2}-\dfrac{1}{2}\right)}}_{\dfrac{1}{A}} \int \prod^{N}_{i=1} \D q_i\, \left[E-V(q^{1}\ldots q^{N})\right]^{\frac{N}{2}-\frac{3}{2}} \Theta\left[E-V(q^1,\ldots,q^N)\right]\;,
\end{equation}
where $\Gamma$ is the Euler Gamma function and $\Theta$ is the Heaviside function.
\\The first four derivatives of $\Omega$ with respect to the total energy, are then
\begin{equation}
\begin{split}
&\Omega'(E)=\left(\dfrac{N}{2}-\dfrac{3}{2}\right)\,\dfrac{1}{A}\, \int  \prod^{N}_{i=1} \D q_i\left(E-V\right)^{\frac{N}{2}-\frac{5}{2}} \Theta\left(E-V\right)\\
&\Omega''(E)=\left(\dfrac{N}{2}-\dfrac{3}{2}\right)\left(\dfrac{N}{2}-\dfrac{5}{2}\right)\,\dfrac{1}{A}\, \int \prod^{N}_{i=1} \D q_i \left(E-V\right)^{\frac{N}{2}-\frac{7}{2}} \Theta\left(E-V\right)\\
&\Omega^{'''}(E)=\left(\dfrac{N}{2}-\dfrac{3}{2}\right)\left(\dfrac{N}{2}-\dfrac{5}{2}\right)\left(\dfrac{N}{2}-\dfrac{7}{2}\right)\,\dfrac{1}{A}\, \int  \prod^{N}_{i=1} \D q_i \big( E-V\big)^{\frac{N}{2}-\frac{9}{2}} \Theta\left(E-V\right)\\ 
&\Omega^{''''}(E)=\left(\dfrac{N}{2}-\dfrac{3}{2}\right)\left(\dfrac{N}{2}-\dfrac{5}{2}\right)\left(\dfrac{N}{2}-\dfrac{7}{2}\right)\left(\dfrac{N}{2}-\dfrac{9}{2}\right)\,\dfrac{1}{A}\, \int \prod^{N}_{i=1} \D q_i \left(E-V_N\right)^{\frac{N}{2}-\frac{11}{2}} \Theta\left(E-V \right)\;,
\end{split}
\end{equation}
where the dependence of the potential on the generalized coordinates $q_{i}$ has been omitted to simplify the notation.
We notice that each derivative of the microcanonical partition function appears divided by $\Omega$. Remembering, as explained in  Ref.\cite{PearsonLaplace1985}, that the microcanonical average $A(q^{1}\ldots q^{N})$ of any function of the generalized coordinates has the form
\begin{equation}
\left \langle A \right\rangle_{\mu c}= \dfrac{1}{\Omega\,A} \,\int \prod^{N}_{i=1} \D q^i A(q^{1}\ldots q^{N}) \big( E-V(q^{1}\ldots q^{N})\big)^{\frac{N}{2}-\frac{3}{2}} \Theta\left(E-V(q^{1}\ldots q^{N})\right)\;,
\end{equation}
and that $\left \langle \left(E-V(q^{1}\ldots q^{N}\right)\right \rangle_{\mu c}=\left \langle K\right \rangle_{\mu c}=N\left \langle \kappa\right \rangle_{\mu c}$, where $\kappa$ is the specific kinetic energy, we obtain for the first four derivatives of the microcanonical partition function
\begin{equation}
\begin{split}
\dfrac{\Omega'}{\Omega}&=\left(\dfrac{N}{2}-\dfrac{3}{2} \right) \left \langle K^{-1} \right \rangle_{\mu c}=\left(\dfrac{1}{2}-\dfrac{3}{2N} \right) \left \langle \kappa^{-1} \right \rangle_{\mu c}\\
\dfrac{\Omega''}{\Omega}&=\left(\dfrac{N}{2}-\dfrac{3}{2} \right) \left( \dfrac{N}{2}-\dfrac{5}{2} \right)\left \langle K^{-2} \right \rangle_{\mu c}=\left(\dfrac{1}{2}-\dfrac{3}{2N} \right) \left( \dfrac{1}{2}-\dfrac{5}{2N} \right)\left \langle \kappa^{-2} \right \rangle_{\mu c}\\ 
\dfrac{\Omega^{'''}}{\Omega}&=\left(\dfrac{N}{2}-\dfrac{3}{2} \right) \left( \dfrac{N}{2}-\dfrac{5}{2} \right)\left( \dfrac{N}{2}-\dfrac{7}{2} \right)\left \langle K^{-3} \right \rangle_{\mu c}=\left(\dfrac{1}{2}-\dfrac{3}{2N} \right) \left( \dfrac{1}{2}-\dfrac{5}{2N} \right)\left( \dfrac{1}{2}-\dfrac{7}{2N} \right)\left \langle \kappa^{-3} \right \rangle_{\mu c}\\ 
\dfrac{\Omega^{''''}}{\Omega}&=\big( \dfrac{N}{2}-\dfrac{3}{2} \big) \big( \dfrac{N}{2}-\dfrac{5}{2} \big)\big( \dfrac{N}{2}-\dfrac{7}{2} \big)\big( \dfrac{N}{2}-\dfrac{9}{2} \big)\big \langle K^{-4} \big \rangle_{\mu c}\\
&=\big(\dfrac{1}{2}-\dfrac{3}{2N} \big) \big( \dfrac{1}{2}-\dfrac{5}{2N} \big)\big( \dfrac{1}{2}-\dfrac{7}{2N} \big)\big( \dfrac{1}{2}-\dfrac{9}{2N} \big)\big \langle \kappa^{-4} \big \rangle_{\mu c}\;.
\end{split}
\end{equation}
In general, it can be verified  that the following expression holds in the case where both \textit{the total energy and the total momentum} are conserved
\begin{equation}
\label{eq:gen_Omegader}
\dfrac{\Omega^{(l)}}{\Omega}=\prod_{m=1}^l\left(\dfrac{1}{2}-\dfrac{2m+1}{2N}\right) \langle \kappa^{-l} \rangle_{\mu c}\;,
\end{equation}
where $(l)$ denotes the first derivate with respect to $E$, whereas in Ref.\cite{PearsonLaplace1985}, where \textit{only the constraint on the total energy is considered}, the previous expression reads
\begin{equation}
\dfrac{\Omega^{(l)}}{\Omega}=\prod_{m=1}^l\left(\dfrac{1}{2}-\dfrac{(2m-1)}{N}\right) \langle \kappa^{-l} \rangle_{\mu c}\;.
\end{equation}
Substituting \eqref{eq:gen_Omegader} in \eqref{eq:derEnt_microcan}, we obtain the derivatives of the microcanonical specific entropy 
as functions of the microcanonical averages of the inverse of the specific kinetic energy $\kappa$,
\begin{equation}
\begin{split}
\dfrac{\partial S}{\partial\varepsilon}=&\left(\dfrac{1}{2}-\dfrac{3}{2N} \right) \left \langle \kappa^{-1} \right \rangle_{\mu c}\\
\dfrac{\partial^{2} S}{\partial\varepsilon^{2}}=& N \left[ \left( \dfrac{1}{2}-\dfrac{3}{2N} \right)\left( \dfrac{1}{2}-\dfrac{5}{2N} \right) \left \langle \kappa^{-2} \right \rangle_{\mu c}-\left( \dfrac{1}{2}-\dfrac{3}{2N} \right)^{2}\left \langle \kappa^{-1} \right \rangle^{2}_{\mu c}\right]\\
\dfrac{\partial^{3} S}{\partial\varepsilon^{3}}=& N^{2}\Bigg[ \left( \dfrac{1}{2}-\dfrac{3}{2N} \right)\left( \dfrac{1}{2}-\dfrac{5}{2N} \right)\left( \dfrac{1}{2}-\dfrac{7}{2N} \right) \left \langle \kappa^{-3} \right \rangle_{\mu c}+\\
&-3\left( \dfrac{1}{2}-\dfrac{3}{2N} \right)^{2}\left( \dfrac{1}{2}-\dfrac{5}{2N} \right)\left\langle \kappa^{-1} \right \rangle_{\mu c} \left \langle \kappa^{-2} \right \rangle_{\mu c}+2\left( \dfrac{1}{2}-\dfrac{3}{2N} \right)^{3}\left\langle \kappa^{-1} \right \rangle^{3}_{\mu c}\bigg] \\
\dfrac{\partial^{4} S}{\partial\varepsilon^{4}}=& N^{3}\bigg[\left( \dfrac{1}{2}-\dfrac{3}{2N} \right)\left( \dfrac{1}{2}-\dfrac{5}{2N} \right)\left( \dfrac{1}{2}-\dfrac{7}{2N} \right)\left(\dfrac{1}{2}-\dfrac{9}{2N} \right) \left \langle \kappa^{-4} \right\rangle_{\mu c}+\\
&-4\left(\dfrac{1}{2}-\dfrac{3}{2N}\right)^{2}\left(\dfrac{1}{2}-\dfrac{5}{2N}\right)\left(\dfrac{1}{2}-\dfrac{7}{2N}\right)\left \langle \kappa^{-1} \right\rangle_{\mu c} \left \langle \kappa^{-3}\right\rangle_{\mu c} +\\
&-3 \left(\dfrac{1}{2}-\dfrac{3}{2N}\right)^{2}\left(\dfrac{1}{2}-\dfrac{5}{2N}\right)^{2} \left \langle \kappa^{-2} \right\rangle^{2}_{\mu_c}+12\left(\dfrac{1}{2}-\dfrac{3}{2N}\right)^{3}\left(\dfrac{1}{2}-\dfrac{5}{2N}\right) \left \langle \kappa^{-2} \right \rangle_{\mu c} \left \langle \kappa^{-1} \right \rangle^{2}_{\mu c}+\\
&-6\left(\dfrac{1}{2}-\dfrac{3}{2N}\right)^{4}\left \langle \kappa^{-1} \right \rangle^{4}_{\mu c} \bigg]\;.
\end{split}
\end{equation}
These derivatives can be used to detect and classify phase transitions tackled in the microcanonical ensemble, as annouced in the Introduction. Moreover, among several other thermodynamic observables we can work out the explicit expressions of two basic observables: the microcanonical temperature 
\begin{equation}
\label{T}
T=\left(\dfrac{\partial S}{\partial \varepsilon } \right)^{-1}=\bigg[\big(\dfrac{1}{2}-\dfrac{3}{2N}\big)\left \langle \kappa^{-1}\right\rangle_{\mu c} \bigg]^{-1}\;,
\end{equation}
which is used to obtain the caloric curve,
and the specific heat
\begin{equation}
\label{c}
c= N \left(\dfrac{\partial T(\varepsilon)}{\partial \varepsilon} \right)^{-1}=- \dfrac{\left( \partial S/ \partial \varepsilon \right)^2}{\left( \partial^2 S/\partial \varepsilon^2 \right)}=\dfrac{1}{N}\,\left[1-\dfrac{\left (1-\dfrac{5}{N}\right)\langle \kappa^{-2}\rangle_{\mu c}}{\left(1-\dfrac{3}{N}\right)\left\langle \kappa^{-1}\right \rangle_{\mu c}^{2}}\right]^{-1}\;.
\end{equation}
In numerical simulations, the microcanonical averages $\langle \cdot\rangle_{\mu c}$ are replaced by time averages of  the same quantities. 
\bigskip

\begin{figure}[H]
 \centering
 \includegraphics[scale=0.35,keepaspectratio=true,angle=0]{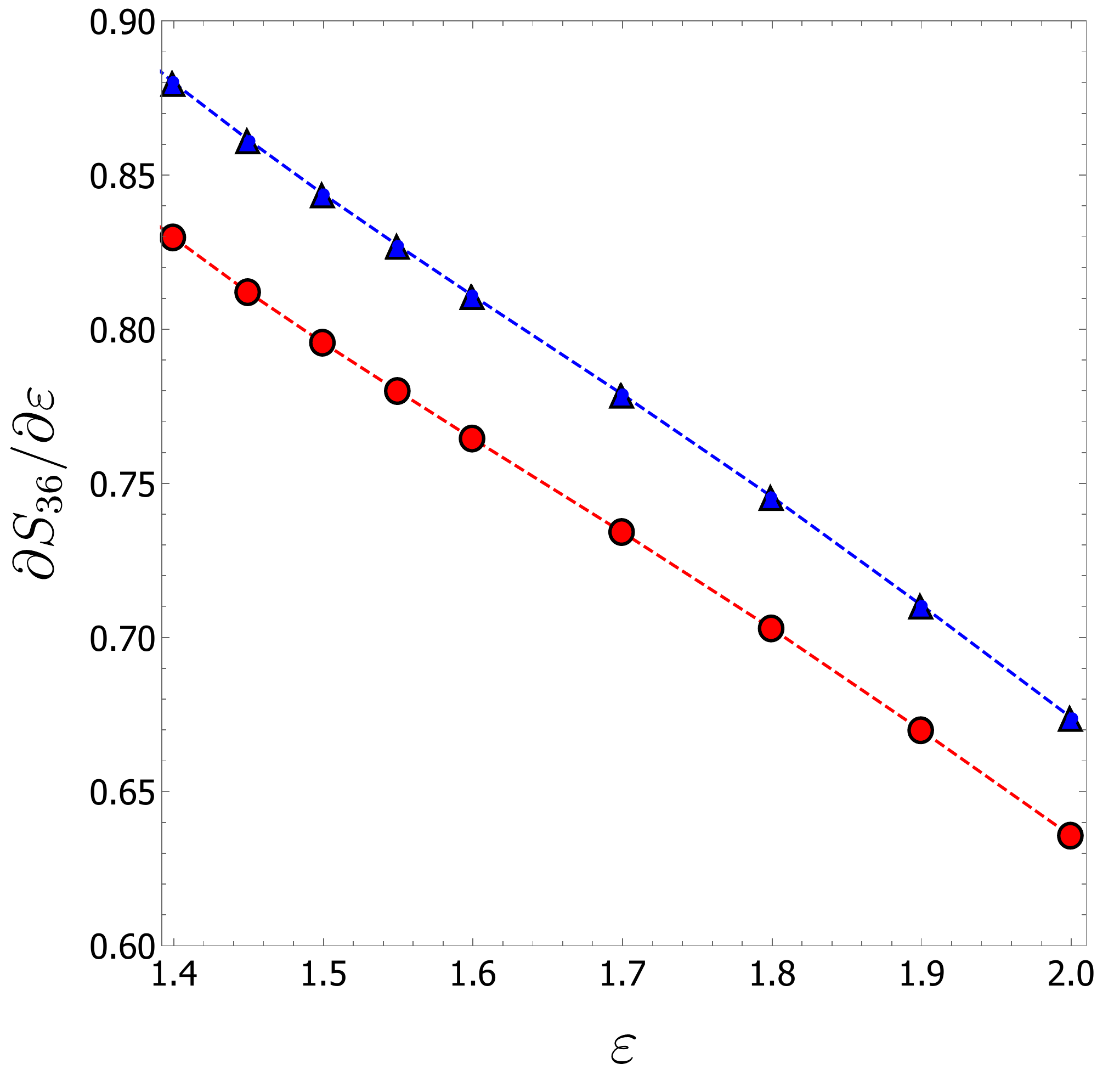} \hskip 1truecm
  \includegraphics[scale=0.35,keepaspectratio=true,angle=0]{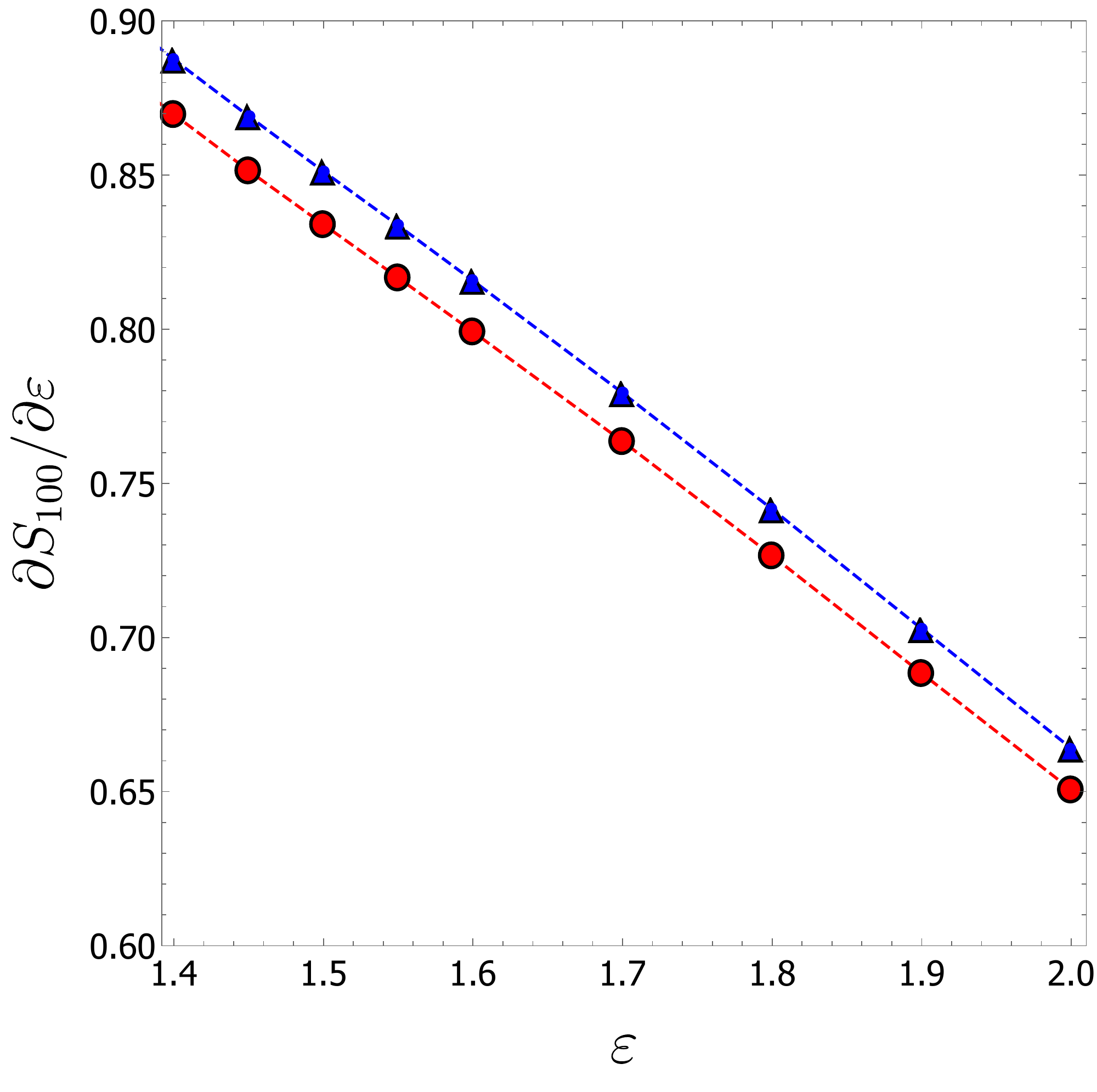}
 \vskip -0.3truecm
 \caption{First derivative with respect to the specific energy $\varepsilon$ at fixed total momentum $\pp = 0$ of microcanonical entropy $S$ for $N=36$ (left panel) and for $N=100$ (right panel). The different markers correspond to different calculation methods: the method using the calculations of this work (red circles) and the original Pearson-Halicioglu formula \cite{PearsonLaplace1985} where the constraint on the 
total momentum was absent (blue triangles). }
\vskip 0.3truecm
\label{entropy}
\end{figure}

In order to provide an evidence for the usefulness of these calculations, we have checked the above given formulas against a specific Hamiltonian system  where the additional conserved quantity, besides the total energy, is the total momentum.
\\The system chosen is the $2$-D XY model, a $2$-vector model on a $2$-dimensional lattice $\lambda$, i.e.  a $2$-dimensional unit-vector $\mathbf{s}_i=(s_{ix},s_{iy})=(\cos\theta_i,\sin\theta_i)$ is associated to each site $i \in \lambda$, and is described by the Hamiltonian 
\begin{equation}
\label{eq:XY_2D_Hamiltonian}
H_{\mathrm{XY}}(\boldsymbol{p},\boldsymbol{\theta})=\sum_{i=1}^{n}\sum_{j=1}^n\dfrac{p_{i,j}^2}{2}-\text{J}\left[2-\cos(\theta_{i,j}-\theta_{i,j+1})-\cos(\theta_{i,j}-\theta_{i+1,j})\right]\,,
\end{equation}
where $J$ represents the coupling constant and $p_i$ are the conjugate momenta to the angles $\theta_i$.
\\According to Noether's theorem, this system has two first integrals, the total energy and, as said before,  the total momentum $\sum\limits_{i\in\lambda}\dot{\theta}$ associated to the global $\text{O}(2)$ symmetry $\theta_i\rightarrow\theta_i+\alpha$.
\bigskip
\\In Figure \ref{entropy} we present the plot of the first derivative of the specific entropy with respect to the specific energy for different lattice sizes: $N=n\times n = 36$ and $N=n\times n = 100$. We notice a significant deviation between the values worked out by means of the Pearson-Halicioglu method (blue symbols and dashed lines) compared to the  values worked out in the present Letter by taking into account the total momentum as additional constant of motion (red symbols and dashed lines). The mismatch is decreasing with $N$ as expected.

\section{Concluding remarks}
In presence of conserved quantities besides the total energy, like total momentum related with invariance under spatial translation or angular momentum related with rotational invariance, one can wonder if the standard microcanonical observables given by the equations reported in Ref.\cite{PearsonLaplace1985} (where only the total energy was taken fixed) can be safely adopted in numerical simulations. As we can see by comparing equations  \eqref{T} and \eqref{c} with the corresponding expressions in Ref.\cite{PearsonLaplace1985}, the corrections asymptotically vanish. Therefore, the computations reported above ensure that for large systems the formulas in Ref.\cite{PearsonLaplace1985} can be safely adopted also in presence of an additional conserved quantity besides the energy.

However, for small $N$ systems, the corrections worked out in the present letter could be of some quantitative relevance. 
\newpage

 %%%%%%%%%%%%%%%%%%%%%%%%%%%%%%%%%%%%%%%%%%%%%%%%%%%%%%%%%%%%%%
% BIBLIOGRAPHY
\bibliography{letter}

\end{document}